\documentstyle[aps,amssymb,psfig,epsf,epsfig,twocolumn]{revtex}
\begin{document}
\newcommand{\ket}[1]{ |  #1  \rangle}
\newcommand{\bra}[1]{ \langle #1   |}
\newcommand{\kett}[1]{ |  #1  \rangle\!\rangle}
\newcommand{\braa}[1]{ \langle \!\langle #1   |}
\newcommand{\scall}[2]{ \langle \!\langle #1   | #2
\rangle\!\rangle}
\newcommand{\scal}[2]{ \langle #1 | #2
\rangle}
\newcommand{\proj}[1]{\ket{#1}\bra{#1}}
\newcommand{\projc}[1]{\ket{#1}_c \,{}_c\bra{#1}}
\newcommand{\proja}[1]{\ket{#1}_a \,{}_a\bra{#1}}
\newcommand{\abs}[1]{ |  #1   |}
\newcommand{\av}[1]{\langle #1\rangle}
\newcommand{\calp}{\mbox{$\cal P$}}
%\tighten
\title{Continuous variable cloning  via network of parametric gates}
\author{Giacomo M. D'Ariano$^{a,b}$, Francesco De Martini$^c$, 
and Massimiliano F. Sacchi$^{a,d}$}
\address{$^a$Theoretical Quantum Optics Group, 
Dipartimento di Fisica `A. Volta' and Istituto Nazionale per la 
Fisica della Materia,\\ Unit\`a di Pavia, Universit\`a di Pavia, via
Bassi 6, I-27100 Pavia, Italy} 
\address{$^b$ Istituto Nazionale di Fisica Nucleare, Gruppo IV,
Sezione di Pavia, via Bassi 6, I-27100 Pavia, Italy}
\address{$^c$Dipartimento di Fisica and Istituto Nazionale per la Fisica della 
Materia,\\ Universit\`{a} di Roma `La Sapienza', Roma 00185, Italy}
\address{$^d$ Optics Section, Blackett Laboratory, Imperial College 
London, London SW7 2BZ, England}
\date{\today}
\maketitle
%%%%%%%%%%%%%%%%%%%%%%%%%%%%%%%%%%%%%%%%%%%%%%%%%%%%%%%%%%%%%%%
\begin{abstract}
We propose an experimental scheme for the cloning machine of 
continuous quantum variables through a network of parametric
amplifiers working as input-output four-port gates.
\end{abstract}
%%%%%%%%%%%%%%%%%%%%%%%%%%%%%%%%%%%%%%%%%%%%%%%%%%%%%%%%%%%%%%%%%%%%%%%%%%%%
%\draft\twocolumn
Since the seminal paper of Bu\v{z}ek and Hillery \cite{buzek96}, much
theoretical work has been done about quantum cloning
\cite{gisin,theor,fort,cerf,clonphase}.  In Ref. \cite{cerf} the problem of
copying the
state of a system with continuous variables has been studied, and a
unitary transformation that clones coherent states with the same
fidelity (equal to $2/3$) 
has been found. However, proposals of its experimental realization
have not appeared yet.  In this Letter, we propose an experimental
scheme to implement such a new kind of cloning. We show that a network
of three parametric amplifiers, each of them working as an
input-output four-port ``gate'', under suitable gain conditions
realizes the one-to-two cloning machine for ``distinguishable''
clones. In fact, optimal cloning machines can be achieved using
parametric down-converters, in two different ways. The first way,
proposed in Ref. \cite{undi}, uses a single ``quantum-injected''
parametric amplifier in a configuration already used in many
experiments \cite{amplent}, where a one-photon state is downconverted
into a many-photons entangled state. This contains clones of the
injected state, which are supported by indistinguishable photons 
\cite{notaref}. This
way can then be used for measurements of permutation-invariant
observables on clones, and allows one to study the clones statistics
\cite{altro}. The second way is the subject of this Letter: a
one-to-two cloning machine for distinguishable clones, based on
parametric gates.  
\par A relevant application of universally covariant cloning is
eavesdropping for quantum cryptography \cite{gisin}. Moreover, 
quantum cloning is of practical interest as a 
tool to engineer novel scheme for joint measurements.  However,
universal covariant cloning is not ideal for such purpose, and a
suitable non universal cloning is needed \cite{jnt}. If one wants to
use quantum cloning to realize joint measurements, cloning must be
optimized for a reduced covariance group, depending on the desired
joint measurement, such that measurements on cloned copies are
equivalent to optimal joint measurements on the original.  
As we will show, this is the case of the cloning map proposed in Ref. 
\cite{cerf}, which is covariant only with respect to the
Weyl-Heisenberg group represented by the displacement operator, 
and which is optimal for the joint measurement of conjugated
quadratures.  
Hence, measures of quality other than fidelity should be used for
optimization, depending on the final use of the output copies. 
This is also indicated by recent studies of 
copying machines designed for information transfer \cite{munr}.  Notice also
that the major problem in quantum teleportation---the Bell
measurement---may need  a general scheme for designing joint
measurements, as shown in Ref. \cite{telep}, 
where the Bell-like measurement is achieved by a
new kind of probability operator-valued measure (POVM) that
generalizes the joint measurements of position-momentum, or the
measurement of the ``direction'' of the angular momentum.  \par We
need to introduce some preliminary mathematics.  Consider the
heterodyne-current operator \cite{yuen} $Z=a+b^\dag$, which satisfies
the commutation relation $[Z,Z^\dag]=0$ and the eigenvalue equation $Z
\kett{z}_{ab}=z\kett{z}_{ab}$, with $z\in\mathbb {C}$.  The
eigenstates $\kett{z}_{ab}$ are given by \cite{zeta1,zeta2}
\begin{eqnarray}
\kett{z}_{ab}\equiv D_a(z)\kett{0}_{ab}=D_b(z^*)\kett{0}_{ab}\;, 
\end{eqnarray}
where $D_d(z)=e^{zd^\dag -z^* d}$ denotes the displacement operator
for mode $d$ and $\kett{0}_{ab}\equiv
\frac{1}{\sqrt\pi}\sum_{n=0}^{\infty}(-1)^n \ket{n}_a\otimes \ket{n}_b$
on the Fock basis.  The eigenstates $\kett{z}_{ab}$ are a complete
orthogonal set with Dirac-delta normalization
${}_{ab}\scall{z}{z'}_{ab}=\delta ^{(2)}(z-z')$.  For $z=0$ the state
$\kett{0}_{ab}$ can be approximated by a physical (normalizable)
state---so-called twin beam---corresponding to the output of a
non-degenerate optical parametric amplifier (NOPA) in the limit of
infinite gain \cite{zeta1}. \par For the following, it is also useful
to evaluate the expression ${}_{cb}\scall{z}{z'}_{ab}$ which is given
by
\begin{eqnarray}
{}_{cb}\scall{z}{z'}_{ab}&=&{}_{cb}\braa{0}
D_c^\dag (z)D_a (z')\kett{0}_{ab}\nonumber \\&= & 
\frac 1\pi D_a(z'){\cal T}_{ac}D_c^\dag (z)
\;,\label{usef}
\end{eqnarray}
where ${\cal T}_{ac}=\sum_n |n\rangle_a \,{}_c\langle n| $ denotes the
{\em transfer} operator \cite{telep} satisfying the relation ${\cal
T}_{ac} \ket{\psi}_c= \ket{\psi}_a$ for any vector $|\psi \rangle
$. Here we briefly transpose the main results of the continuous
variable cloning in Ref. \cite{cerf}, in a compact formalism suited to
the following treatment. 
The input state at the cloning machine can be written
\begin{eqnarray}
\ket{\phi}=\ket{\varphi}_c\otimes \int_ {\mathbb C} d^2 z\,f(z,z^*)
\kett{z}_{ab}\;, 
\end{eqnarray}
where $\ket{\varphi}_c$ is the original in the Hilbert space ${\cal
H}_c$, to be cloned in ${\cal H}_c$ itself and ${\cal H}_a$, whereas
${\cal H}_b$ is an ancillary Hilbert space.  We do not specify for the
moment the explicit form of the function $f(z,z^*)$.  The cloning
transformation is realized by the unitary operator \cite{cerf}
\begin{eqnarray}
U=\exp\left[\left(X_c+i Y_c \right)Z^\dag -
\left(X_c -i Y_c\right)Z\right]\;,\label{uk}
\end{eqnarray}
with $X_c$ and $Y_c$ denoting the conjugated quadratures for mode $c$,
namely $X_c=(c+c^\dag )/2$ and $Y_c=(c-c^\dag)/2i$. Notice that one
has $U \kett{z}_{ab}=D_c^{\dag}(z)\,\kett{z}_{ab}$. The state after
the cloning transformation is given by $\ket{\phi}_{out}=U\ket{\phi}$.
Let us evaluate the one-site restricted density matrix $\varrho_c$ and
$\varrho_a$ corresponding to the state $\ket{\phi}_{out}$, for the
Hilbert spaces ${\cal H}_c$ and ${\cal H}_a$ supporting the clones.
For $\varrho_c$ one has
\begin{eqnarray}
\varrho_c&=&\hbox{Tr}_{ab}[\proj{{\phi }_{out}}] \nonumber \\&=& 
\int_{\mathbb C}d^2w\int_{\mathbb C}d^2z\int_{\mathbb C}d^2z'\,
f(z,z^*)f^*(z',z'^*)\nonumber \\&\times& 
{}_{ab}\braa{w}D_c^{\dag}(z) \projc{\varphi}D_c(z')\otimes 
\kett{z}_{ab}\,{}_{ab}{\scall{z'}{w}}_{ab}\nonumber
\\&= & \int_{\mathbb C}d^2z |f(z,z^*)|^2 D_c^{\dag}(z)\proj{\varphi}
D_c(z)\;,\label{cl1}
\end{eqnarray}
where we have evaluated the trace by using completeness and 
orthogonality of the eigenstates $\kett{w}_{ab}$ of $Z$.  
For $\varrho_a$, using Eq. (\ref{usef}), one has
\begin{eqnarray}
\varrho_a&=&\hbox{Tr}_{cb}[\proj{{\phi}_{out}}] \nonumber \\&=& 
\int_{\mathbb C}d^2w\int_{\mathbb C}\frac{d^2z}{\pi}\int_{\mathbb C}
\frac{d^2z'}{\pi}\,
f(z,z^*)f^*(z',z'^*)
\nonumber \\&\times  & D_a(z ){\cal T}_{ac}\left[ 
D_c^{\dag}(w)D_c^{\dag}(z)
\projc{\varphi}D_c(z')D_c(w)\right]\nonumber \\&\times & 
{\cal T}_{ca}D_a^\dag (z')\nonumber \\&= & 
\int_{\mathbb C} d^2 w\, 
| \stackrel {\sim }{f} (w,w^{*})|^2 D_a^{\dag }(w)
\proja{\varphi}D_a(w) \;,\label{after}
\end{eqnarray}
where $\stackrel{\sim }{f}(w,w^{*})$ denotes the Fourier transform
over the complex plane
$\stackrel{\sim }{f}(w,w^*)=\int_{\mathbb C}
\frac{d^2z}{\pi}\,e^{wz^*-w^*z}\,f(z,z^*)$.
Hence, for $f(z,z^*)=\sqrt{2/\pi}\,e^{-|z|^2}$ one has 
two identical clones $\varrho_c=\varrho_a$, which are given 
by the original state $|\varphi \rangle $ degraded 
by Gaussian noise. The state preparation $\ket{\chi}$ 
pertaining to the Hilbert space 
${\cal H}_a\otimes {\cal H}_b$ is given by \cite{jnt}
\begin{eqnarray}
&&\ket{\chi}=
\sqrt{\frac 2\pi}\int_{\mathbb C}
d^2z\,e^{-|z|^2} \kett{z}_{ab} = \frac{2\sqrt2}{3} \times 
\label{twb}\\
&&\sum_{n=0}^{\infty}\left(-\frac
13\right)^n \ket{n}_a\otimes\ket{n}_b
%\nonumber \\&= & 
=e^{\scriptsize{\hbox{atanh}}\frac 13(ab-a^\dag b^\dag )}
\ket{n}_a\otimes\ket{n}_b
\nonumber 
\;.
\end{eqnarray}
One recognizes in Eq. (\ref{twb}), the twin-beam state at the output
of a NOPA with total number of photons $N=\bra{\chi}a^\dag a+ b^\dag b
\ket{\chi}=1/4$.
\par Quantum cloning allows one to engineer new joint measurements. In
fact, suitable measurements on the cloned copies are equivalent to a
joint measurement on the original.  Let us now consider a joint
position-momentum on the original copy through the present
scheme. More precisely, in our case measuring two quadratures on the
two clones will be equivalent to the joint measurement of a couple of
conjugated quadratures on the original, namely to a
heterodyne measurement. This can be shown as follows. Consider the
entangled state $\varrho $ at the output of the cloning machine, after
tracing over the ancillary mode $b$. One has
\begin{eqnarray}
\varrho&=&
\hbox{Tr}_{b}[\proj{{\phi }_{out}}] \nonumber \\&=& 
\frac 12 P_{ca}(\projc {\varphi}\otimes \openone _a) P_{ca}
\;,\label{trc}
\end{eqnarray}
where $P_{ca}$ is the projector given by 
\begin{eqnarray}
P_{ca}=V (\projc{0}\otimes {\openone }_a) V^\dag \;,\label{vv}
\end{eqnarray}
with $V=\exp [\frac \pi 4 (c^\dag a-c a^\dag )]$. Measuring the
quadratures $X_c$ and $Y_a$ is then equivalent to perform the
measurement on the original state $\ket{\varphi }_c$ described by the
POVM
\begin{eqnarray}
F(x,y)&=&\hbox{Tr} _{a}[P_{ca}
\projc{x} \otimes \proja{y} P_{ca}]
\;, 
\end{eqnarray}
where $\ket{x}_c$ and $\ket{y}_a$ denote the eigenstates of  $X_c$ and
$Y_a$, respectively. From the following relations \cite{zeta2}
\begin{eqnarray}
&&V^\dag \projc{x} \otimes \proja{y} V \nonumber \\&&= 
2\kett{\sqrt 2 (x-iy)}_{ca}\,{}_{ca} \braa{\sqrt2 (x-iy)}
\;,\nonumber \\& & 
V \ket{\alpha }_c\otimes \ket{\beta }_a =
\ket{(\alpha +\beta )/\sqrt 2 }_c
\otimes\ket{(\beta -\alpha )/\sqrt 2}_a \;,\nonumber \\& & 
{}_c \langle 0 \kett{z}_{ca}=\frac {1}{\sqrt \pi} \ket{z^*}_a\;,\label{foll}
\end{eqnarray}
one obtains
\begin{eqnarray}
F(x,y)=\frac 1 \pi \projc{x+iy}\;,\label{coh}
\end{eqnarray}
namely the coherent-state POVM, which is the well-known optimal 
joint measurement of conjugated quadratures $X_c$ and $Y_c$ [In
Eq. (\ref{coh}) and in the last two lines of Eq. (\ref{foll}) single-mode 
vectors are coherent states].
\par\noindent Eqs. (\ref{trc}) and (\ref{vv}) allows one to show that
the cloning machine here considered is covariant with respect to the
Weyl-Heisenberg group, represented by the displacement operator. 
One has 
\begin{eqnarray}
&&\frac 12 P_{ca}(D_c(\alpha )\projc {\varphi}D^\dag _c(\alpha )
\otimes \openone _a) P_{ca} \nonumber \\&&= 
D_c(\alpha )\otimes D_a(\alpha ) 
\,\varrho\,D_c^\dag (\alpha )\otimes D^\dag _a(\alpha ) 
\;.\label{}
\end{eqnarray}
\par
\begin{figure}[hbt]
\begin{center}
\epsfxsize=.7 \hsize\leavevmode\epsffile{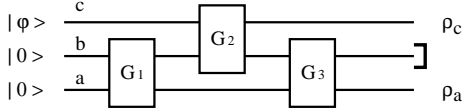}
\end{center}
\caption{Network of parametric amplifiers, each of them working as an
input-output four-port gate, to realize the continuous variable
one-to-two cloning. The values of the gain parameters are given in
Eqs. (\ref{g3}).} \label{f:clongate}\end{figure}

 In the following we show that the unitary evolution in
Eq. (\ref{uk}) can be obtained from a network of three NOPA's 
under suitable gain conditions. 
We rewrite Eq. (\ref{uk}) as $U=\exp(B+A)$, 
with $B=ca^\dag -c^\dag a$ and $A=bc-b^\dag c^\dag$. Upon defining 
$C=ab-a^\dag b^\dag$, one easily checks the 
commutation relations $[C,A]=B$, $[C,B]=A$, and $[B,A]=C$. Hence, 
the following identity holds 
\begin{eqnarray}
e^{\lambda C}\,A\,e^{-\lambda C}=\cosh(\lambda )A+\sinh (\lambda
)B\;.\label{aca}
\end{eqnarray}
From Eq. (\ref{aca}) one obtains the realization for the
operator $U$
\begin{eqnarray}
U=\lim_{\lambda \rightarrow\infty}e^{\lambda C}\,e^{2e^{-\lambda
}A}\,e^{-\lambda C}\;.\label{lim}
\end{eqnarray}
Each term in the product of the r.h.s of Eq. (\ref{lim}) is realized
by a NOPA. The continuous variable cloning from one to two copies is
then achievable in the limit $\lambda \rightarrow\infty$ through the 
network of parametric amplifiers depicted in
Fig. \ref{f:clongate}.  Notice that the evolution operator for the
generation of the input state of Eq. (\ref{twb}) can be absorbed into 
the last factor of the product in Eq. (\ref{lim}), yielding the
overall unitary transformation
\begin{eqnarray}
U'=e^{\lambda C}\,e^{2e^{-\lambda
}A}\,e^{(\scriptsize{\hbox{atanh}}\frac 13-\lambda )C}\;.\label{lim2}
\end{eqnarray} 
The gain values of the three amplifiers are constrained as follows
\begin{eqnarray}
&&G_1=\hbox{cosh}^2(\lambda -\hbox{atanh }1/3)\;,\nonumber \\ 
&&G_2=\hbox{cosh}^2(2e^{-\lambda} )\;, \nonumber \\
&&G_3=\hbox{cosh}^2 \lambda \;. 
\label{g3}
\end{eqnarray}
Notice that a cascade of $N$ networks could produce $2^N$
clones. However, as shown in Ref. \cite{fort}, this is not an
efficient way to produce multiple clones. 
\par In the following we provide the POVM corresponding to 
the measurement of two 
quadratures $X_c(\varphi) =(c^\dag e^{i\varphi}+c
e^{-i\varphi})$ and $X_a(\theta) =(a^\dag e^{i\theta }+a
e^{-i\theta})$ over the two clones, in the case of realistic cloning
($\lambda < \infty $).  For $0<\theta-\varphi <\pi$, one obtains the 
POVM
\begin{eqnarray}
F_\lambda (x,x';\varphi, \theta)
&=&\frac 1{4\pi} \,\frac {C |\delta |^2}{\sqrt{CD -E^2}}
\nonumber \\&\times &
S^\dag(\xi)\,D(\alpha \,\delta )\,\varrho \,D^\dag (\alpha\,
\delta )S(\xi)\; 
\end{eqnarray}
where 
\begin{eqnarray}
\alpha &=&-\frac i2 x+\frac {Cx'-Ex}{2\sqrt{CD-E^2}}\;, \nonumber \\
\varrho &=&\frac {4}{C |\delta |^2+2}\left(\frac{C|\delta |^2-2}
{C|\delta |^2+2}\right)^{c^\dag c} \;,\nonumber \\ C&=&(\sinh
\varepsilon \,\sinh \lambda ')^2+\frac 12 \sinh ^2 \varepsilon
\nonumber \\ D&=&(\cosh \lambda \,\cosh \lambda '-\sinh \lambda
\,\cosh \varepsilon \,\sinh \lambda ')^2 \nonumber \\ &+&\frac
12(\sinh \lambda \,\sinh \varepsilon )^2-\frac 12 \nonumber
\\E&=&\cos(\varphi-\theta)[\sinh \varepsilon \,\sinh \lambda' 
(\cosh \lambda \,\cosh \lambda ' \nonumber \\  &-& \sinh
\lambda \,\cosh \varepsilon \,\sinh \lambda ')-\frac12 \cosh
\varepsilon \,\sinh \lambda \,\sinh \varepsilon ]\;,
\nonumber
\\\delta &=&-(|\gamma |^2-|\beta 
|^2)^{-1/2}\,e^{-i\hbox{\scriptsize arg}\gamma }\, \nonumber \\
\beta &=&\frac 14 \cosh \varepsilon \,e^{i\varphi}\left(i+\frac
{E}{\sqrt{CD-E^2}}\right) \nonumber \\& +&\frac 14\sinh \lambda
\,\sinh \varepsilon \, e^{i\theta}\,\frac
{C}{\sqrt{CD-E^2}}\;,\nonumber \\ \gamma &=&\frac 14 \cosh
\varepsilon \,e^{i\varphi}\left(-i+\frac {E}{\sqrt{CD-E^2}}\right)
\nonumber \\& +&\frac 14\sinh \lambda \,\sinh \varepsilon \,
e^{i\theta}\,\frac {C}{\sqrt{CD-E^2}}\;,\nonumber \\
 \xi&=&\hbox{acosh}(|\gamma \delta |)\,e^{i(\hbox{\scriptsize arg}\gamma
+ \hbox{\scriptsize arg}\beta )}\, 
\end{eqnarray}
with $\varepsilon =-2e^{-\lambda }$ and $\lambda '=\lambda
-{\hbox{atanh}}\frac 13$. 
\par Notice that for $\lambda \rightarrow \infty $
and $\theta -\varphi=\pi/2$ one gets the result in Eq. (\ref{coh}), 
namely one achieves the ideal POVM for simultaneous measurement of
conjugated quadratures.  
\par Now we can evaluate the added noise for simultaneous measurement
of conjugated quadratures over the two clones. One has 
the following input-output relations between the expectation
values over the two clones $ \langle \cdots \rangle _o$ at the output
and the same expectation $ \langle \cdots \rangle _i$ for the
original copy at the input 
\begin{eqnarray}
&&\left\langle X_c\right\rangle _o =\cosh \varepsilon
\left\langle X_c\right\rangle _i \;,\nonumber \\& &
\left\langle Y_a\right\rangle _o =\sinh \lambda
\sinh \varepsilon \left\langle Y_c\right\rangle _i
\;,\nonumber \\& & \left\langle X_c ^2\right\rangle _o
=\cosh ^2 \varepsilon \left\langle X_c^2
\right\rangle _i\nonumber \\&+& \frac 14 \sinh^2 \varepsilon (2\sinh ^2
\lambda '+1) \;,\nonumber \\& & \left\langle Y_a^2
\right\rangle _o =\sinh ^2\lambda \sinh ^2 \varepsilon \left\langle
Y_c^2\right\rangle _i\nonumber \\&+&\frac 14
(\cosh \lambda \cosh \lambda '-\sinh \lambda \sinh \lambda ' \cosh
\varepsilon )^2 \nonumber \\&+ & \frac 14 (\sinh \lambda \cosh \lambda
'\cosh \varepsilon -\cosh \lambda \sinh \lambda ' ) ^2 \;. 
\end{eqnarray}
\par\noindent In the limit of  $\lambda \rightarrow\infty$ one has 
\begin{eqnarray}
&&\langle \Delta X_c^2\rangle _o\rightarrow 
\langle \Delta X_c^2\rangle _i +\frac 14\;, \nonumber \\& & 
\langle \Delta Y_a^2\rangle _o\rightarrow 
\langle \Delta Y_c^2\rangle _i +\frac 14\;,
\end{eqnarray}
which proves the optimality of the joint measurement \cite{yuen2}.
\par
\begin{figure}[hbt]
\begin{center}
\epsfxsize=.6 \hsize\leavevmode\epsffile{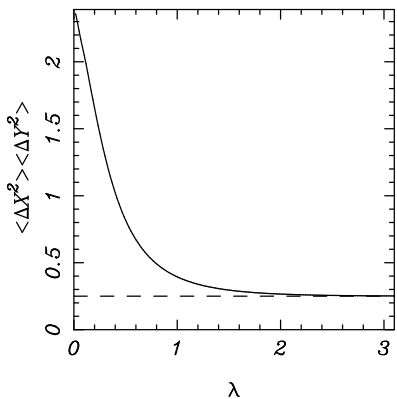}
\end{center}
\caption{Optimal joint measurement of conjugated quadratures via
continuous variable cloning and homodyne detection, for arbitrary
coherent input state.  The bound $\langle \Delta X ^2 \rangle \langle
\Delta Y^2 \rangle =1/4$ is achieved by increasing the parameter
$\lambda $ which provides the optimal cloning for $\lambda \rightarrow
\infty $ (the plot is independent of the amplitude of the coherent
state).}
\label{f:added}\end{figure}

 The behavior of the product of variances for the simultaneous
measurement of $X_c$ and $Y_c$ via homodyne detection over clones is
plotted in Fig. \ref{f:added}, for arbitrary coherent state (for
which $\langle \Delta X_c^2\rangle _i =\langle \Delta Y_c^2\rangle
_i=1/4$). Notice that for increasing value of $\lambda $ the
optimality of the joint measurement is rapidly achieved.
\par The condition $f(z,z^*)=\stackrel{\sim }{f}(z,z^*)$ 
for Eqs. (\ref{cl1}) and (\ref{after}) in order to
obtain identical clones can be
satisfied also by a bivariate Gaussian of the form
\begin{eqnarray}
f(z,z^*)=\sqrt{ 2\over \pi}\,\exp\left(-\frac{\hbox{Re}^2z}{\sigma ^2}
-\sigma ^2\,\hbox{Im}^2 z\right)\;.\label{sig}
\end{eqnarray}
In such case, as shown in Ref. \cite{jnt}, the cloning 
transformation becomes optimal for the joint measurement of
noncommuting quadratures $X_\phi$ and $X_{-\phi}$, at angles which
depend on the parameter $\sigma $ in Eq. (\ref{sig}) as
$\phi=\hbox{arctg}(\sigma ^2)$.
\par As regards the experimental realization of the network in Fig. 1,
the scheme which is presently engineered in our lab in Rome works with
an injection method similar to the one used in the implementation of
the all-optical Schroedinger-cat of Ref.  \cite{amplent}. Three
identical equally oriented nonlinear crystals of beta-barium-borate
cut for Type II phase matching are excited by coherent beams derived
from a common UV beam at wavelength $\lambda _{p}=400nm$. In the
present experiment the UV beam is supplied by second harmonic
generation of the output of a Coherent MIRA TI:SA mode-locked laser
consisting of a train of 150 fs pulses emitted at a rate of 76
MHz. The average output power does not exceed 0.6 W, and then the
amplification gain is of the order $g\simeq 0.02$, where
$G=\hbox{cosh}^2 g$ in Eqs. (\ref{g3}).  We expect a far larger
efficiency by the forthcoming implementation within the apparatus of a
regenerative NOPA Coherent REGA9000. In this case the value 
of $g$ is multiplied by an adjustable factor 
in the range $10-50$, and the cloning 
efficiency is expected to increase by the same factor. The nonlinear
crystals emit $\mathbf{\pi}$-entangled photons with wavelength
$\lambda =800nm$ over two modes determined by two fixed $1mm$ pinholes
placed 2 meters away from the source crystals.  
The conditions imposed by Eqs. (\ref{g3}) are achieved by a precise 
setting of the intensity of the three single-mode UV pump beams by use of 
adjustable Circular Neutral Density Filters Newport 946. 
Great care is taken in space-mode filtering
which select the injection modes through the pinholes. The
spatio-temporal superposition for such short pulses and mode matching
at homodyne detectors are the main experimental challenges. 
\par In conclusion, our experimental scheme based on a network of
three NOPA's is designed for engineering quantum clones of harmonic
oscillator states. For increasing gain, optimality is achieved in
performing joint quadrature measurements via cloning. 
\par\noindent {\em Acknowledgments.} 
This work has been supported by the Italian Ministero 
dell'Universit\`a e della Ricerca Scientifica e Tecnologica (MURST)
 under the co-sponsored project 1999
``Quantum Information Transmission And Processing: Quantum
Teleportation And Error Correction''. M.F.S. acknowledges the
Leverhulme Trust Foundation for partial support.

\end{document}